\documentclass{article}
\usepackage{amsthm}
\usepackage{amsmath}
\usepackage{amssymb}
\usepackage{newalg}
\usepackage{graphicx}

\newcommand{\set}[1]{\left\{{#1}\right\}}

\DeclareMathOperator{\wt}{wt}

\theoremstyle{plain}
\newtheorem{sample}{Example}

\title{Polar Codes with Dynamic Frozen Symbols and Their Decoding by Directed Search}
\author{Peter Trifonov and Vera Miloslavskaya
}

\begin{document}
\sloppy
\maketitle
\begin{abstract}
A novel construction of polar codes with dynamic frozen symbols is 
proposed. The proposed codes are subcodes of extended BCH codes, which 
ensure sufficiently high minimum distance. Furthermore, a decoding  algorithm is proposed, which 
employs estimates of the not-yet-processed bit channel error 
probabilities to perform directed search in code tree, reducing thus the 
total number of iterations. 
\end{abstract}
\section{Introduction}    
Polar codes were recently shown to be able to achieve the capacity of a 
wide class of communication channels \cite{arikan2009channel}. However,
the performance of polar codes of moderate length appears to be quite poor.
This is both due to suboptimality of  the successive cancellation (SC) decoding algorithm and low minimum distance
of polar codes. The first problem was addressed in \cite{tal2011list}, where a list decoding algorithm for polar codes was introduced.
Similar stack-based decoding algorithm was presented in \cite{niu2012crcaided}.  To solve the problem of low minimum distance, serial 
concatenation with an outer CRC code, i.e. taking some subcode of the original polar code,  was suggested in \cite{tal2011list}. 
Numeric results show that this approach significantly improves the performance, although no non-trivial estimates of the minimum 
distance of the obtained codes are available.  

Observe that pre-encoding the data with CRC introduces dependencies between information symbols of the inner polar code.
In this paper this idea is generalized by constructing these dependencies in such way, so that the obtained code is a subcode of another code with sufficiently high minimum distance. Furthermore, a novel decoding algorithm for polar codes
is derived, which performs directed search in code tree for the most probable codeword.

The paper is organized as follows. Section \ref{sBackground} presents the 
background on polar codes and SC decoding. 
Polar codes with dynamic frozen symbols are introduced in Section 
\ref{sSubcodes}. A novel decoding algorithm for polar codes is derived  in Section 
\ref{sImprovedStack}. Numeric results  illustrating the performance of the proposed codes and improved decoding algorithm
are provided in Section \ref{sNumeric}. Finally, some conclusions are drawn.

\section{Background}
\label{sBackground}
\subsection{Polar codes and the successive cancellation algorithm}
$(n=2^m,k)$ polar code is a linear block code generated by $k$ rows of 
matrix $A=B_mF^{\otimes m}$, where $B_m$ is the bit-reversal permutation 
matrix, $F=\begin{pmatrix}1&0\\1&1\end{pmatrix}$, and $\otimes m$ denotes 
$m$-times Kronecker product of the matrix with itself \cite{arikan2009channel}. The particular rows 
to be used in a generator matrix are selected so that the decoding error 
probability is minimized. Hence, a codeword of a classical polar code is obtained as 
$c=uA$, where $u_i=0, i\in \mathcal F,$ and $\mathcal F\subset\set{0,\ldots,n-1}$ is the 
set of $n-k$ frozen bit subchannel indices.  Observe that $AA=I$. Hence, 
the parity check matrix of a polar code is given by rows of $A^T$ with 
indices in $\mathcal F$.

The SC decoding algorithm at phase $i$
computes 
$P(u_0^i|y_0^{n-1})=\frac{P(y_0^{n-1},u_0^{i-1}|u_i)}{2P(y_0^{n-1})}, 
u_i\in \set{0,1},$
where $a_s^t=(a_s,\ldots,a_t)$, $y_0,\ldots,y_{n-1}$ are the noisy symbols 
obtained by transmitting codeword symbols $c_0,\ldots,c_{n-1}$ over a 
binary input memoryless output-symmetric channel. The decoder makes 
decision 
\begin{equation}
\label{mSCDecisionRule}
\widehat{u}_i=\begin{cases} \arg \max_{u_i\in\set{0,1}} P(y_0^{n-1},u_0^{i-1}|u_i),& i\not \in \mathcal F \\0,&\text{otherwise}.\end{cases}
\end{equation}
 This decision 
is used at subsequent steps instead of the true value of $u_i$ to determine the values of 
$u_{i+1},\ldots,u_{n-1}$.  It was shown in \cite{arikan2009channel} that 
these calculations can be implemented with complexity $O(n\log n)$. 

It is possible to re-formulate the above described algorithm in terms of 
log-likelihood ratios $L_i=\log 
\frac{P(y_0^{n-1},u_0^{i-1}|u_i=1)}{P(y_0^{n-1},u_0^{i-1}|u_i=0)}$, and 
compute their probability distributions assuming that zero codeword is 
transmitted, and all previous estimates $u_0,\ldots, u_{i-1}$ are correct. 
This can be implemented via density evolution \cite{tal2011how} or its 
Gaussian approximation \cite{chung2001analysis,trifonov2012efficient}.  Then one can compute 
the probabilities $p_i$ of incorrect estimation of each $u_i$, and construct $\mathcal F$ as 
the set of $n-k$ indices $i$ with the largest $p_i$. The minimum distance 
of  the obtained polar code is given by $2^{t_0}$, where 
$t_0=\min\set{\wt(i)|i\in \set{0,\ldots,n-1}\setminus \mathcal F}$, and 
$\wt(i)$ denotes the number of $1$'s in the binary expansion of integer $i$. 
Observe that this method for construction of   polar codes is not 
guaranteed to be optimal if some other decoding algorithm (e.g. list or stack SC) is used.

\subsection{Stack SC decoding}
The main problem with the SC decoding algorithm is 
that it cannot recover  errors occuring at its early stages. Since at the 
$i$-th phase the decoder takes into account only a subset of rows of 
the parity check matrix of the polar code corresponding to frozen symbols 
$i': i'<i, i'\in \mathcal F$, the decisions performed at early stages are 
quite unreliable, and may need to be revised as soon as additional 
parity check constraints are taken into accout.  This problem can be 
avoided by keeping a list of most likely paths. Each path is identified by vector $u_0^{i-1}$
of the values of already processed symbols $u_j$.  The list may include 
either $L$ paths of the same length \cite{tal2011list}, or a varying 
number of paths of different lengths arranged in a stack (in fact, 
priority queue) \cite{niu2012crcaided}.   If $i\in \mathcal F$, then the 
path can be extended to $(u_0,\ldots,u_{i-1},0)$. Otherwise, two possible 
extensions $(u_0,\ldots,u_{i-1},0)$ and $(u_0,\ldots,u_{i-1},1)$ need to 
be considered.   At each iteration stack decoder selects for 
extension the path $u_0^{i-1}$ with the largest value of 
\begin{equation}
\label{mNiuMetric}
M(u_0^{i})=\begin{cases}
M(u_0^{i-1}),& i\in \mathcal F,\\
\log P(u_0^{i}|y_0^{n-1}),&i \not \in \mathcal F.
\end{cases}
\end{equation}
This can be considered as an instance of the Dijkstra algorithm for 
finding the shortest (meaning most likely in the context of decoding) path in a graph (code tree).

\section{Polar codes with dynamic frozen symbols}
\label{sSubcodes}
It appears that polar codes constructed using density evolution method  have 
quite small minimum distance. To solve this problem, observe that it is not 
necessary to set $u_i=0, i\in \mathcal F$. The $i$-th frozen symbol can be equal to 
any pre-defined function of non-frozen symbols $u_j, j<i$. This does not 
affect the behaviour of the SC decoder and performance of bit 
subchannels induced by the linear transformation given by matrix $A$.

Observe that $n\times n$ matrix $A$ is invertible. This implies that any $(n=2^m,k,d)$ linear code $C$ 
with check matrix $H$
can be obtained as an appropriate subspace of its row space. Namely, information vector $u$ results in
a polar codeword being also a codeword of $C$ if $$u\underbrace{AH^T}_{V^T}=0.$$
Let $i_j=\max \set{t\in \set{0,\ldots,n-1}| V_{j,t}=1}, 0\leq j<n-k$. By applying elementary row operations to 
matrix $H$, it is possible to obtain $V$ such that for any $t\in 
\set{0,\ldots,2^m-1}$ there exists at most one $j: i_j=t$. Let $\mathcal 
F=\set{t|\exists j: i_j=t}$.  Let $S_j=\set{t|V_{j,t}=1, t<i_j}$. Hence, 
one obtains the dynamic freezing constraints on information symbols of a 
polar code 
\begin{equation}
u_j=\sum_{t\in S_j} u_t, j\in \mathcal F.
\end{equation}
In the case of $S_j=\emptyset$ one obtains classic static frozen symbols.
This enables one to perform decoding of any binary block code using the SC decoder or any of its variations. 
Namely, \eqref{mSCDecisionRule} can be replaced with 
\begin{equation}
\label{mSCDecisionRuleDynamic}
\widehat{u}_i=\begin{cases} \arg \max_{u_i\in\set{0,1}} P(y_0^{n-1},u_0^{i-1}|u_i),& i\not \in \mathcal F \\\sum_{t\in S_i} u_t,&i\in \mathcal F.\end{cases}
\end{equation}
However, the set of non-frozen bit subchannels obtained for a generic linear binary code using the above described
method includes, in general, many bad subchannels, while many good bit subchannels are frozen. This causes the 
SC decoder performance to be much worse 
compared to state-of-the-art decoding algorithms \cite{Valembois2004box}. 

It was shown in \cite{kolesnik1968cyclic,kasami1968new,delsarte1970generalized} that 
a punctured Reed-Muller code of order $r$ and length $2^m$ contains a 
subcode equivalent to the cyclic code 
 with generator polynomial  $g(x)$ having roots $\alpha^i: 1\leq 
\wt(i)<m-r, 1\leq i\leq 2^m-2$, where $\alpha$ is a primitive element of $GF(2^m)$.
On the other hand, $(r,m)$ Reed-Muller code can be considered as a special case of a polar code with $\mathcal F=\set{i\in \set{0,\ldots,2^m-1}|\wt(i)<m-r}$.
Hence, given an extended BCH code, one can identify an appropriate Reed-Muller supercode, so that all bit 
subchannels $i$ with sufficiently small $\wt(i)$ are frozen. In general, such bit subchannels have high error probability under SC decoding.
 
\begin{sample}
\label{ex1676}
Consider $(16,7,6)$ extended BCH code. The generator polynomial of the corresponding non-extended code has roots $\alpha, \alpha^3$ and their conjugates, where $\alpha$ is a primitive root 
of $x^4+x+1$.
The constraints on 
vector $u$, such that $u\mathcal A$ is a permuted codeword of this code, 
are given by
\begin{equation*}
u\begin{tiny}\left(
\setlength{\arraycolsep}{1pt}
\begin{array}{cccccccccccccccc}
1&0&0&0&0&0&0&0&0&0&0&0&0&0&0&0\\    
1&0&0&0&0&0&0&0&1&0&0&0&0&0&0&0\\    
1&0&0&0&1&0&0&0&0&0&0&0&0&0&0&0\\    
1&0&0&0&1&0&0&0&1&0&0&0&1&0&0&0\\    
1&0&1&0&0&0&0&0&0&0&0&0&0&0&0&0\\    
1&0&1&0&0&0&0&0&1&0&1&0&0&0&0&0\\    
1&0&1&0&1&0&1&0&0&0&0&0&0&0&0&0\\    
1&0&1&0&1&0&1&0&1&0&1&0&1&0&1&0\\    
1&1&0&0&0&0&0&0&0&0&0&0&0&0&0&0\\    
1&1&0&0&0&0&0&0&1&1&0&0&0&0&0&0\\    
1&1&0&0&1&1&0&0&0&0&0&0&0&0&0&0\\    
1&1&0&0&1&1&0&0&1&1&0&0&1&1&0&0\\    
1&1&1&1&0&0&0&0&0&0&0&0&0&0&0&0\\    
1&1&1&1&0&0&0&0&1&1&1&1&0&0&0&0\\    
1&1&1&1&1&1&1&1&0&0&0&0&0&0&0&0\\    
1&1&1&1&1&1&1&1&1&1&1&1&1&1&1&1\\    
\end{array}
\right)
\setlength{\arraycolsep}{1pt}
\begin{pmatrix}
1+a+a^2+a^3&a^2+a^3&1\\
a+a^2+a^3&a^3      &1\\
1+a^2+a^3&a+a^3    &1\\
a^2+a^3&a^3        &1\\
1+a+a^3&a^2+a^3    &1\\
a+a^3&1+a+a^2+a^3  &1\\
1+a^3&1+a+a^2+a^3  &1\\
a^3&a+a^3          &1\\
1+a+a^2&1          &1\\
a+a^2&1            &1\\
1+a^2&a+a^3        &1\\
a^2&a^2+a^3        &1\\
1+a&1+a+a^2+a^3    &1\\
a&a^3              &1\\
1&1                &1\\
0&0                &1\\
\end{pmatrix}\end{tiny}=0.
\end{equation*}
%\left(
%\setlength{\arraycolsep}{1pt}
%\begin{array}{cccccccccccccccc}
%\end{array}\right)\end{tiny}=0
%1+a+a^2+a^3&a+a^2+a^3&1+a^2+a^3&a^2+a^3&1+a+a^3&a+a^3&1+a^3&a^3&1+a+a^2&a+a^2&1+a^2&a^2&1+a&a&1&0\\ 
%a^2+a^3&a^3&a+a^3&a^3&a^2+a^3&1+a+a^2+a^3&1+a+a^2+a^3&a+a^3)&1&1&a+a^3&a^2+a^3&1+a+a^2+a^3&a^3&1&0\\
%1&0&0&0&0&0&0&0&0&0&0&0&0&0&0&0
Multiplying matrices, expanding their elements in the standard basis and 
applying elementary column operations, one obtains
\begin{equation}
u
\left(\setlength{\arraycolsep}{1pt}
\begin{array}{cccccccccccccccc}
0&0&0&0&0&1&0&0&0&0&1&0&1&0&0&0\\
0&0&0&0&0&0&0&0&0&1&1&0&0&0&0&0\\
0&0&0&0&0&1&1&0&0&1&0&0&0&0&0&0\\
0&0&0&0&0&0&0&0&1&0&0&0&0&0&0&0\\
0&0&0&1&0&1&0&0&0&0&0&0&0&0&0&0\\
0&0&0&0&1&0&0&0&0&0&0&0&0&0&0&0\\
0&0&1&0&0&0&0&0&0&0&0&0&0&0&0&0\\
0&1&0&0&0&0&0&0&0&0&0&0&0&0&0&0\\
1&0&0&0&0&0&0&0&0&0&0&0&0&0&0&0
\end{array}
\right)^T=0
\end{equation}
This means that $u_0=u_1=u_2=u_4=u_8=0$ (static frozen symbols), and 
$u_5=u_3$, $u_9=u_5+u_6$, $u_{10}=u_9$, $u_{12}=u_5+u_{10}=u_6$ (dynamic frozen 
symbols). $u_3,u_6, u_7,u_{11},u_{13},u_{14}, u_{15}$ are
non-frozen symbols.
\end{sample}
Unfortunately, exploiting the relationship of e-BCH and Reed-Muller codes
is not sufficient to exclude all bad bit subchannels from the set of non-frozen ones.
The set of dynamic frozen subchannels of low-rate e-BCH 
codes includes many good ones, while a lot of subchannels with high error 
probability remain unfrozen. List SC decoding with  extremely large list size
has to be used in order to obtain the performance comparable with other decoding algorithms.   

To avoid this problem and obtain a $(2^m,k,\geq d)$ code suitable for use with SC decoder, we propose to 
construct the information symbol constraints (i.e. identify dynamic frozen symbols) for a 
high-rate $(2^m,k',d)$ e-BCH code with sufficiently high minimum distance $d$, 
and additionally freeze $k'-k$ bit subchannels with highest error probability, as determined by density evolution.  
\begin{sample}
Let us construct a $(16,6,6)$ code based on $(16,7,6)$ e-BCH code 
considered in Example \ref{ex1676}, by optimizing it for the case of binary erasure channel with erasure probability $p_{0,0}=0.5$. 
The bit subchannel Bhattacharyya parameters are given by 
\cite{arikan2009channel}
\begin{eqnarray*}
p_{m,2j}&=&2p_{m-1,j}-p_{m-1,j}^2\\
p_{m,2j+1}&=&p_{m-1,j}^2.
\end{eqnarray*}
Hence, one obtains
 $p_4=(
\underline{0.9999}, \underline{0.992}, \underline{0.985}, 0.77,  
\underline{0.96}$, $\underline{ 0.65}, 0.53, 0.1,             
\underline{0.9},   \underline{0.47},  \underline{0.35}$,  $3.7\cdot 10^{-2}, \underline{0.23},            1.5\cdot 10^{-2}, 
7.8\cdot 10^{-3}, 1.5\cdot 10^{-5})$.  Here the values 
corresponding to frozen bit subchannels of the e-BCH code are underlined. 
It can be seen that $u_3$ has the largest erasure probability $0.77$, and has to 
be frozen to obtain the required code.
\end{sample}
\begin{sample}
Consider construction of a $(1024,512)$ code.  There exists a 
$(1024,513,116)$ e-BCH code, which cannot, however, be decoded 
efficiently with (list) SC decoder. On the other hand, pure polar code 
optimized for AWGN channel with $E_b/N_0=2 dB$ has minimum distance 16.  
One can take a $(1024,913,24)$ e-BCH code and freeze $401$ additional bit 
subchannels to obtain a $(1024,512,\geq 24)$ polar code with dynamic 
frozen symbols. 
\end{sample}
%It was shown in \cite{hussami2009performance} that the minimum distance of  pure polar codes is upper bounded by $d\leq 2 ^{\frac{m}{2}+c\sqrt{m}}$, where $c$ depends on code rate. 

\section{Decoding with directed search}
\label{sImprovedStack}
MAP decoding of a polar code requires finding a sequence of 
information bits $u_0^{n-1}$, such that $P(u_0^{n-1}|y_0^{n-1})$ is maximizes, subject 
to information symbol freezing constraints. 
In the case of SC decoding one obtains 
\begin{equation}
\label{mProbFact}
P(u_0^{n-1}|y_0^{n-1})=P(u_0^{i-1}|y_0^{n-1})\prod_{j=i}^{n-1}P(u_j|u_0^{j-1},y_0^{n-1}).
\end{equation}
The original SC decoding algorithm \cite{arikan2009channel} operates locally on code tree, and 
selects at each phase $i\not\in \mathcal F$ the most likely value $u_i$, appending it to the path being reconstructed. List SC decoding algorithm keeps 
$L$ paths $u_0^{i-1}$, extends them with both values of $u_i$ and eliminates least likely paths.
Stack SC decoding algorithm \cite{niu2012crcaided}  keeps a set $\mathcal U$ of path of variable length, and at each iteration 
selects for extension the path having  most probable head $u_0^{i-1}$. $\mathcal U$ contains initially an empty path.

The genie stack SC decoder should select at each iteration the path $u_0^{i-1}$ which maximizes \eqref{mProbFact}. In this case, exactly $n$
iterations would be performed. However, in a real decoder for a branch 
$u_0^{i-1}$ in a code tree only $P(u_0^{i-1}|y_0^{n-1})$ is available. 
This causes the decoder to switch frequently between different paths, 
increasing thus the number of iterations needed to find the most probable codeword.
To avoid this problem, we propose to replace the second term in \eqref{mProbFact} with its expected value over all possible received vectors $y_0^{n-1}$.
Assume for the sake of simplicity that zero codeword has been transmitted.
Then for the  correct  path $u_0^{i-1}=\mathbf 0^i=(0,\ldots,0)$
one obtains $E[P(u_0^{n-1}=\mathbf 0^{n-1}|y_0^{n-1})]=P(u_0^{i-1}|y_0^{n-1})\cdot\prod_{j=i}^{n-1}E[P(u_j=0|u_0^{j-1}=\mathbf 0^j,y_0^{n-1})]=P(u_0^{i-1}|y_0^{n-1})\phi(i),$ 
where 
\begin{equation}
\label{mHeuristic}
\phi(i)=\prod_{j=i}^{n-1}(1-P_j),
\end{equation}
and $P_j$ is  the $j$-th subchannel error probability, provided that exact values of all previous bits $u_i, i<j$, are available. For a given channel, $P_j$ can be pre-computed using  density evolution. At each iteration the decoder should select for extension a path $u_0^{i-1}$, such that $P(u_0^{i-1}|y_0^{n-1})\cdot \phi(i)$ is maximized.

However, it may happen that $\phi(i)$ is less than the actual value of $\prod_{j=i}^{n-1}P(u_j|u_0^{j-1},y_0^{n-1})$.
In this case the decoder may proceed along an incorrect path. If there exists a codeword $c: P(c_0^{n-1}|y_0^{n-1})>P(u_0^{i-1}|y_0^{n-1})\phi(i)$ for some $i$,
this results in decoding error, i.e. the performance of the proposed 
algorithm may be worse than that of a maximum-likelihood decoder. 
%This can be avoided by constructing $\phi(i)$ so that $\phi(i)>\prod_{j=i}^{n-1}P(u_j|u_0^{j-1},y_0^{n-1})$ with sufficiently high probability $\Pi$.  The decoding error probability under this algorithm (assuming infinite memory) can be estimated as $P_A\leq P_{ML}+k(1-\Pi)$, where $P_{ML}$ is the maximum-likelihood decoding error probability, and $k$ is the number of non-frozen bit subchannels. 
This approach can be considered  as an instance of $A$-algorithm (see \cite{sorokine1998sequential} and references therein), with $\phi(i)$ being a heuristic function, estimating the cost of the unexplored part of the paths.

\begin{figure}
\begin{algorithm}{Decode}{y_0^{n-1},L,C}
\CALL {push}(1,\epsilon); N\=1; q\=(0,\ldots,0)\\
\begin{WHILE}{true}
U\=\CALL{PopMax}();N\=N-1\\
i\=|U|; q_i\=q_i+1\\
\begin{IF}{i=n}
\RETURN U
\end{IF}\\
\begin{IF}{i\in \mathcal F}
u\=\sum_{t\in S_i}u_t;
S\=P(U.u|y_0^{n-1})\\
\CALL {push}(S\cdot\phi(i+1),U.u); N\=N+1
\ELSE
\begin{WHILE}{N>C-2}
\CALL{KillPath}(\CALL{PopMin}());N\=N-1
\end{WHILE}\\
S_0\=P(U.0|y_0^{n-1}); S_1\=P(U.1|y_0^{n-1})\\
\CALL {push}(S_0\cdot\phi(i+1),U.0); N\=N+1\\
\CALL {push}(S_1\cdot\phi(i+1),U.1); N\=N+1
\end{IF}\\
\begin{IF}{q_i\geq L}
\begin{FOR}{\text{each path $U$ in the priority queue}}
\begin{IF}{|U|<L}
\CALL{Unqueue}(U)\\
\CALL{KillPath}(U); N\=N-1
\end{IF}
\end{FOR}
\end{IF}
\end{WHILE}
\end{algorithm}
\caption{Decoding with directed search}
\label{fStackDecoder}
\end{figure}
\begin{figure*}
\centering \includegraphics[width=0.75\textwidth]{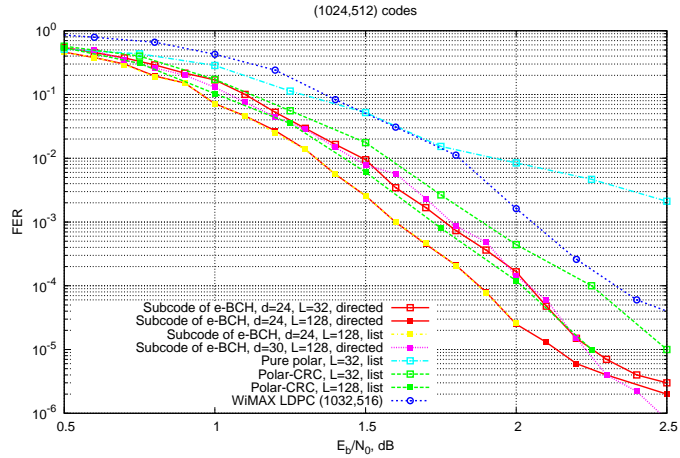}
\caption{Performance of length $1024$ codes}
\label{f1024Perf}
\end{figure*}

One should keep the number of paths tracked by the decoder limited.
Various techniques to 
do this were considered in \cite{niu2012crcaided}, in particular:
\begin{itemize}
\item  Let $q_i$ be the number of times a path of length $i$ was 
extracted from the priority queue. If $q_i\geq L$, all paths shorter than 
$i+1$ should be dropped from the queue.  That is, $L$ is the maximal list 
size at each decoding phase.
\item If the number of paths $N$ in the priority queue exceeds its capacity $C$, 
the least probable ones should be dropped.
\end{itemize}

In practice, the probability of correct path being dropped as a result of application of these rules
significantly exceeds the probability of an incorrect path being extended till phase $n$ due to $\phi(i)$ being less than the actual probability of its tail.
Figure \ref{fStackDecoder} summarizes the proposed algorithm. The algorithm is presented in probability domain, but it can be easily modified to work with logarithms of path probabilities.
Obviously, this algorithm is applicable to classical polar codes as well.

The algorithm makes use of functions $Push(S,U)$, $PopMax()$ and $PopMin()$, which push 
into the priority queue path $U$ with score $S$, and extract from it the paths with 
the highest and smallest scores, respectively. $\epsilon$ denotes an empty vector, $U=(u_0,u_1,\ldots)$ is a vector of information symbols corresponding to a path,  and   $U.a$ denotes a vector obtained by
appending value $a$ to vector $U$.  Function $KillPath(U)$ corresponds to dropping a path, while line 17 corresponds
to cloning a path. Efficient implementation of these operations was given 
in \cite{tal2011list}.  Observe that the values of $u_i$ can be obtained
from the data structures maintained by Tal-Vardy list decoding algorithm.
Function $Unqueue$ removes an element from the priority queue.     
A priority queue supporting the required operations  can be implemented using a red-black tree \cite{Cormen2001introduction}.

\section{Numeric results}
\label{sNumeric}
Figure \ref{f1024Perf} presents the performance of  $(1024,512)$ 
polar codes with dynamic frozen symbols obtained as subcodes of 
$(1024,913,24)$  and $(1024,883,30)$ e-BCH codes, conventional polar code, a polar code 
concatenated with outer CRC one \cite{tal2011list} constructed for $E_b/N_0=2 dB$, and a WiMAX LDPC code. 
Decoding of polar codes was performed using the proposed directed search algorithm.
For comparison, the performance of SC list decoder is also shown. 
It appears that the proposed directed search algorithm provides  
exactly the same performance as list decoding algorithm with the same list size $L$, so the results for the latter algorithm are reported only for the case of $L=128$.

It can be seen that the improved minimum distance of the proposed codes results in substantially 
better performance compared to pure polar codes. The code with design minimum 
distance $d=24$ outperforms both polar-CRC and LDPC ones. 
  However, as in the case of 
polar-CRC codes, large list size $L$ is needed to fully exploit the 
error-correcting capability of the proposed codes.  Furthermore, increasing 
design minimum distance causes many good bit subchannels to be frozen, 
which results in performance degradation of SC list/stack decoder in the 
low-SNR region. However, at high SNR larger minimum distance enables one 
to avoid error floor.  
%For the case of $d=30$ all decoding errors were non-ML ones, so even better performance can be obtained by increasing $L$.

\begin{figure}
\includegraphics[width=0.5\textwidth]{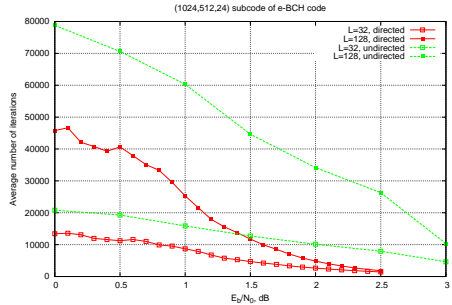}
\caption{Average number of iterations}
\label{fItCount}
\end{figure}

Figure \ref{fItCount} presents average number of  iterations performed by 
the SC stack decoding algorithm (see Figure 
\ref{fStackDecoder}) with and without the proposed 
directed search method. In the latter case (i.e. with $\phi(i)=1$)  the 
algorithm reduces to the one presented in \cite{niu2012crcaided}. It can 
be seen that employing directed search dramatically reduces the number of 
iterations compared to the original stack algorithm, especially in  the 
high SNR region. 

\begin{figure}
\includegraphics[width=0.5\textwidth]{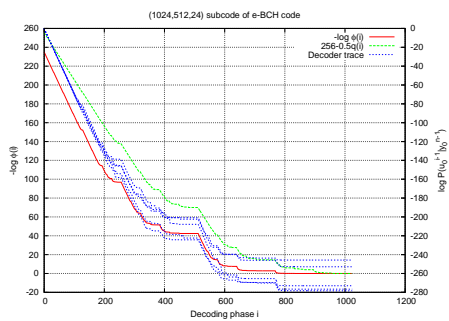}
\caption{Heuristic function behaviour}
\label{fHeuristic}
\end{figure}

Figure \ref{fHeuristic} illustrates the behaviour of logarithmic heuristic 
function $-\log \phi(i)$ (see \eqref{mHeuristic}), obtained with Gaussian approximation for density evolution,
as well as a number of decoder traces, which correspond to sequences of values $\log P(u_0^{i-1}|y_0^{n-1})$ for correct paths. Furthermore, appropriately scaled number of not-yet-processed frozen 
symbols $q(i)=|\mathcal F\cap \set{i,\ldots,n-1}|$ at phase $i$ is shown. 
If $\phi(i)$ were an exact value of the probability of the unexplored part of the correct path at phase $i$,
$\log P(u_0^{i-1}|y_0^{n-1})-(-\log \phi(i))$ would remain a constant. However, it can be seen that for small $i$
$\phi(i)$ overestimates this probability, while for large $i$ the score of the correct path $u_0^{i-1}$
 may vary significantly around its initial estimate $\phi(0)$. Nevertheless, as it was shown above, the performance of the 
proposed decoding algorithm appears to be exactly the same as the list decoder with the same list size $L$, which does not employ any heuristic functions.

It can be also seen that the decoder trace closely follows $q(i)$, i.e. correct path probability drops mostly while 
processing frozen symbols. Observe that the path score function \eqref{mNiuMetric}, which was used in \cite{niu2012crcaided} 
to implement undirected search for the correct path in a code tree, exhibits the opposite behaviour.

\section{Conclusions}
In this paper a novel construction of polar codes with dynamic frozen 
symbols was proposed. It enables one to ensure that the code has 
sufficiently high minimum distance by employing an appropriate e-BCH supercode. 
However, increasing the design minimum distance of the e-BCH supercode  
results in many good bit subchannels to be frozen. This may cause the 
SC decoder to take wrong path at some step. Avoding this problem requires 
one to perform list decoding with sufficiently large list size. In order 
to keep the decoding complexity low, one has to keep the design minimum 
distance of polar codes many times smaller compared to the one achievable
with pure BCH or related codes. This implies that development of more 
advanced decoding algorithms for polar codes may result in further 
performance improvements.

Another contribution of this paper is a decoding algorithm for polar codes with or without dynamic frozen symbols, 
which employs an estimate of bit subchannel error probabilities 
to select the most likely path  to be extended. This significantly reduces 
the number of iterations performed by the decoder without any practical performance loss. 
\section*{Acknowledgements}
This work was supported by Samsung Electronics, and partially by Russian Foundation for Basic Research under the grant 12-01-00365-a.
\bibliographystyle{ieeetran}
\bibliography{coding,comm,trifonov,math}

% Generated by IEEEtran.bst, version: 1.13 (2008/09/30)
\begin{thebibliography}{10}
\providecommand{\url}[1]{#1}
\csname url@samestyle\endcsname
\providecommand{\newblock}{\relax}
\providecommand{\bibinfo}[2]{#2}
\providecommand{\BIBentrySTDinterwordspacing}{\spaceskip=0pt\relax}
\providecommand{\BIBentryALTinterwordstretchfactor}{4}
\providecommand{\BIBentryALTinterwordspacing}{\spaceskip=\fontdimen2\font plus
\BIBentryALTinterwordstretchfactor\fontdimen3\font minus
  \fontdimen4\font\relax}
\providecommand{\BIBforeignlanguage}[2]{{%
\expandafter\ifx\csname l@#1\endcsname\relax
\typeout{** WARNING: IEEEtran.bst: No hyphenation pattern has been}%
\typeout{** loaded for the language `#1'. Using the pattern for}%
\typeout{** the default language instead.}%
\else
\language=\csname l@#1\endcsname
\fi
#2}}
\providecommand{\BIBdecl}{\relax}
\BIBdecl

\bibitem{arikan2009channel}
E.~Arikan, ``Channel polarization: A method for constructing capacity-achieving
  codes for symmetric binary-input memoryless channels,'' \emph{IEEE
  Transactions On Information Theory}, vol.~55, no.~7, pp. 3051--3073, July
  2009.

\bibitem{tal2011list}
I.~Tal and A.~Vardy, ``List decoding of polar codes,'' in \emph{Proceedings of
  IEEE International Symposium on Information Theory}, 2011.

\bibitem{niu2012crcaided}
K.~Niu and K.~Chen, ``{CRC}-aided decoding of polar codes,'' \emph{IEEE
  Communications Letters}, vol.~16, no.~10, October 2012.

\bibitem{tal2011how}
I.~Tal and A.~Vardy, ``How to construct polar codes,'' \emph{IEEE Transactions
  On Information Theory}, 2011, submitted for publication.

\bibitem{chung2001analysis}
S.-Y. Chung, T.~J. Richardson, and R.~L. Urbanke, ``Analysis of sum-product
  decoding of low-density parity-check codes using a {Gaussian}
  approximation,'' \emph{IEEE Transactions on Information Theory}, vol.~47,
  no.~2, February 2001.

\bibitem{trifonov2012efficient}
P.~Trifonov, ``Efficient design and decoding of polar codes,'' \emph{IEEE
  Transactions on Communications}, vol.~60, no.~11, pp. 3221 -- 3227, November
  2012.

\bibitem{Valembois2004box}
A.~Valembois and M.~Fossorier, ``Box and match techniques applied to
  soft-decision decoding,'' \emph{IEEE Transactions on Information Theory},
  vol.~50, no.~5, May 2004.

\bibitem{kolesnik1968cyclic}
V.~Kolesnik and E.~Mironchikov, ``Cyclic {Reed-Muller} codes and their
  decoding,'' \emph{Problems of Information Transmission}, vol.~4, no.~4, pp.
  15--19, 1968.

\bibitem{kasami1968new}
T.~Kasami, S.~Lin, and W.~Peterson, ``New generalizations of the {Reed-Muller}
  codes part i: Primitive codes,'' \emph{IEEE Transactions on Information
  Theory}, vol.~14, no.~2, March 1968.

\bibitem{delsarte1970generalized}
P.~Delsarte, J.~Goethals, and F.~MacWilliams, ``On generalized {Reed-Muller}
  codes and their relatives,'' \emph{Information and control}, vol.~16, pp.
  403--442, 1970.

\bibitem{sorokine1998sequential}
V.~Sorokine and F.~Kschischang, ``A sequential decoder for linear blockcodes
  with a variable bias-term metric,'' \emph{IEEE Transactions On Information
  Theory}, vol.~44, no.~1, January 1998.

\bibitem{Cormen2001introduction}
T.~H. Cormen, C.~E. Leiserson, R.~L. Rivest, and C.~Stein, \emph{Introduction
  to Algorithms}, 2nd~ed.\hskip 1em plus 0.5em minus 0.4em\relax The MIT Press,
  2001.

\end{thebibliography}
\end{document}